\documentclass[twocolumn]{aastex6}

\usepackage{float}

\fullcollaborationName{}

\begin{document}


\title{Ultra Violet Escape Fractions from Giant Molecular Clouds During Early Cluster Formation}
\shorttitle{UV Escape Fractions From GMCs}


\author{Corey Howard}
\affil{Department of Physics and Astronomy, McMaster University \\
1280 Main St.~W, Hamilton, ON L8S 4M1, Canada}

\author{Ralph Pudritz\altaffilmark{1,2,3}}
\affil{Department of Physics and Astronomy, McMaster University \\
1280 Main St.~W, Hamilton, ON L8S 4M1, Canada}

\and

\author{Ralf Klessen\altaffilmark{4}}
\affil{Zentrum f\"ur Astronomie der Universit\"at Heidelberg, Institut f\"ur Theoretische Astrophysik \\
Albert-Ueberle-Str. 2, 69120 Heidelberg, Germany}


\altaffiltext{1}{Origins Institute, McMaster University, 1280 Main St.~W, Hamilton, ON L8S 4M1, Canada}
\altaffiltext{2}{Zentrum f\"ur Astronomie der Universit\"at Heidelberg, Institut f\"ur Theoretische Astrophysik, Albert-Ueberle-Str. 2, 69120 Heidelberg, Germany}
\altaffiltext{3}{Max-Planck Institute f\"ur Astronomie, K\"onigstuhl 17, 69117 Heidelberg, Germany}
\altaffiltext{4}{Interdisziplin\"ares Zentrum f\"ur Wissenschaftliches Rechnen der Universit\"at Heidelberg}

\begin{abstract}

The UV photon escape fraction from molecular clouds is a key parameter for understanding the ionization of the Interstellar Medium (ISM), 
and extragalactic processes, such as cosmic reionization. We present the ionizing photon flux and the corresponding photon escape fraction (f$_{esc}$) arising as a consequence of star cluster formation in a turbulent, 
10$^6$ M$_{\odot}$ GMC, simulated using the code FLASH. We make use of sink particles to represent young, star-forming clusters coupled with a radiative transfer scheme to calculate the emergent UV flux. We find that the ionizing photon flux across the cloud boundary is highly variable in 
time and space due to the turbulent nature of the intervening gas. The escaping photon fraction remains at $\sim$5\% for the first 
2.5 Myr, followed by two pronounced peaks at 3.25 and 3.8 Myr with a maximum f$_{esc}$ of 30\% and 37\%, respectively. 
These peaks are due to the formation of large HII regions, that expand into regions of lower density and some of which
reach the cloud surface. However, these phases are short lived and f$_{esc}$ drops sharply as the HII regions are
quenched by the central cluster passing through high-density material due to the turbulent nature of the cloud. We find an average f$_{esc}$ of 15\% with factor of two variations over 1 Myr timescales. Our
results suggest that assuming a single value for f$_{esc}$ from a molecular cloud is in general a poor approximation, and
that the dynamical evolution of the system leads to large temporal variation.
\end{abstract}




\section{Introduction} \label{sec:intro}

The escape of UV photons from massive stars in young star clusters within molecular clouds drives many critical processes in the Interstellar and Intergalactic Medium. The radiation released by stars contributes to the Interstellar Radiation Field (ISRF) which has the highest energy densities at optical and UV wavelengths \citep{Draine2011},
 the strength of which was first estimated by \cite{Habing} to be $\sim$ 4$\times$10$^{-14}$ erg cm${-3}$ for 12.4 eV photons. Later authors have further characterized 
the strength of the UV portion of the ISRF by including wavelength dependence \citep{Draine1978,Mathis}.

The ISRF, and its interactions with gas and dust, is responsible for determining the chemical, thermal, and ionization state of the Interstellar Medium (ISM) via photoionization, photodissociation, photoelectric 
heating, and absorption and re-emission by dust grains \citep{Draine2011}. Since most UV photons are generated by massive stars in the range 10-100 M$_{\odot}$, they
contribute significantly to the strength of the ISFR and significantly alter the state of the ISM in their vicinity, even when considering their short lifetimes.

It has also become clear in recent years that UV ionizing photons from galaxies hosting Active Galactic Nuclei (AGN) are not sufficient to completely reionize the Intergalactic medium (IGM) by z$=$6 
\citep{Fan2006, Robertson2013}. Instead, fainter dwarf galaxies, with masses as low as $\sim$10$^8$ M$_{\odot}$, are needed to provide the remaining UV photons via their stellar content \citep{Wise2014,Xu2015}.
These low mass galaxies may contribute up to $\sim$40\% of the total ionizing photons required for reionization \citep{Wise2014}. 

In order to contribute to reionization, ionizing photons produced in these galaxies must escape into the intergalactic medium (IGM) \citep{Robertson2010}. The exact fraction of photons, f$_{esc}$, which 
escape their host galaxies, however, is a debated topic. For bright, high redshift galaxies, measured via the Lyman continuum, f$_{esc}\sim$7\% \citep{Siana2015} 
but this number can be as high as $\sim$30\% for fainter Lyman-$\alpha$ emitting galaxies \citep{Nestor2013}. Estimates of f$_{esc}$ from the Large Magellanic Cloud (LMC) and the Small Magellanic Cloud (SMC) based on HII region mapping suggest global escape fractions of 4\% and 11\% respectively \citep{Pelle2015}.

Simulations which attempt to quantify f$_{esc}$ for both
high and low mass galaxies have been performed, but these results often vary by orders of magnitude. For example, \cite{Pard2011} found f$_{esc}$ $<$1\% for high
redshift dwarf galaxies, while later numerical works have found f$_{esc}$ $>$10\% \citep{Raz2010,Ferrera2013,Pard2015}. Moreover, f$_{esc}$ can vary by orders of 
magnitude over the lifetime of the galaxy \citep{Pard2011}.

As illustrated by the numerical simulations in \cite{Pard2011}, the distribution of dense gas in star forming regions is one of the main constraints on f$_{esc}$ from a galaxy. This suggests that detailed modeling 
  of f$_{esc}$ from dense regions \emph{within} galaxies is required to fully understand the trends observed in more global simulations. Giant molecular clouds (GMC) are the densest regions of the galactic ISM, and 
they are the sites where all known star formation takes place. Studying the escape of UV photons from GMCs is therefore also important for a better understanding of cosmological reionization.

The GMC environment is complex, consiting of filaments produced by supersonic turbulence out of which stars, and clusters, ultimately form \citep{Bertoldi1992,Lada2003,MacLow2004,McKee2007,Andre2014,Klessen2016}. Stars which form in this environment can then
alter their surroundings via the emission of radiation, producing HII regions. The complexity of this problem necessitates the use of numerical simulations. While simulations 
of GMCs which include star formation and radiative transfer have been completed \citep{Dale2005,Murray2010,Peters2010,Krumholz2010,Bate2012,Klassen2012-2,Walch2013}, these studies do not examine the fraction of photons that escape the cloud.

In this paper, we address the critical question of UV escape fractions from turbulent molecular clouds by computing f$_{esc}$ from 10$^6$ M$_{\odot}$ GMCs. We employ our suite of simulations which simulated star cluster formation and radiative feedback within young, 10$^6$ M$_{\odot}$ GMCs which have varying initial virial parameters
 \citep{Howard2016}. We model the early evolution of star clusters, defined here as less than 5 Myr, since the effects of supernovae are not included.  We found that, despite producing large HII regions, the inclusion of radiative feedback only suppressed the formation of clusters by a few percent. In comparison,
varying the initial virial parameter from 0.5 to 5 (ie. bound to unbound) reduced the efficiency of cluster formation by $\sim$34\%. The high final star formation efficiencies (SFEs), 
which range from 18\% to 34\%, suggest that radiative feedback alone is not responsible for limiting star formation but that initially unbound clouds better reproduce locally
observed GMCs. 

Given that we have computed the structure and dynamics of cluster forming clouds undergoing radiative feedback, we can now address the question of what fraction of the UV photons produced by the massive stars in clusters escapes the molecular cloud.

We present maps of the ionizing photon flux escaping the cloud to demontrate its highly nonuniform nature in space. We also present f$_{esc}$ (used hereafter to represent the escape fraction from a GMC) during the first 4 Myr of the GMC's evolution which
are shown to be highly variable in time and peaks at $\sim$35\% with a long term average value of $\sim$15\%. The variable nature of f$_{esc}$ is attributed to HII regions which dramatically vary in both shape and size due to the dynamical nature of the gas and embedded clusters.

\section{Method} \label{sec:method}

Below, we provide a brief description of our numerical methods and subgrid model for star cluster formation.

We have simulated a 10$^6$ M$_{\odot}$ GMC using the Adaptive Mesh Refinement (AMR) code FLASH \citep{FLASH} which includes self-gravity, radiative transfer, star cluster formation,
and cooling processes (see \cite{Howard2016} for more detail). This cloud mass was chosen in particular because high mass GMCs contain most of the molecular mass in the Milky Way and are host to the most massive stellar clusters \citep{MacLow2004,McKee2007,Klessen2016}.

The cloud is initially overlaid with a turbulent velocity field which is composed of a mixture of solenoidal and compressive turbulence with a Burgers spectrum (as in \cite{Girichidis2011}).
We selected a configuration with an initial virial paramater of 3, corresponding to an initial Mach number of 73. We chose
this simulation in particular out of the suite presented in \cite{Howard2016} because we found that initially unbound clouds best reproduce the properties of massive GMCs in the Milky Way. The radius of the cloud is
33.8 pc. The initial average density of the GMC is $n$ $=$ 100 cm$^{-3}$, with a density profile which is uniform in the inner half of the cloud and 
decreases as r$^{-3/2}$ in the outer half.

The package PARAMESH is used for the adaptive mesh portion of FLASH \cite{FLASH}. The grid is refined at locations with sharp density or temperature contrasts
to improve the resolution near filaments and HII regions. The minimum cell size in our simulation is 0.13 pc.

To model gas cooling, we employ the method from \citet{Banerjee2006} which treats cooling via molecular line emission, gas-dust interactions, H$_2$ dissociation, and radiative diffusion 
in the optically thick limit. The cooling rates from \citet{Neufeld} are used to treat molecular line emission, while the treatment in \citet{Goldsmith} cools the gas via 
gas-dust transfer.

\begin{figure*}
\centering
\begin{tabular}{ c c }
  \includegraphics[width=0.45\textwidth]{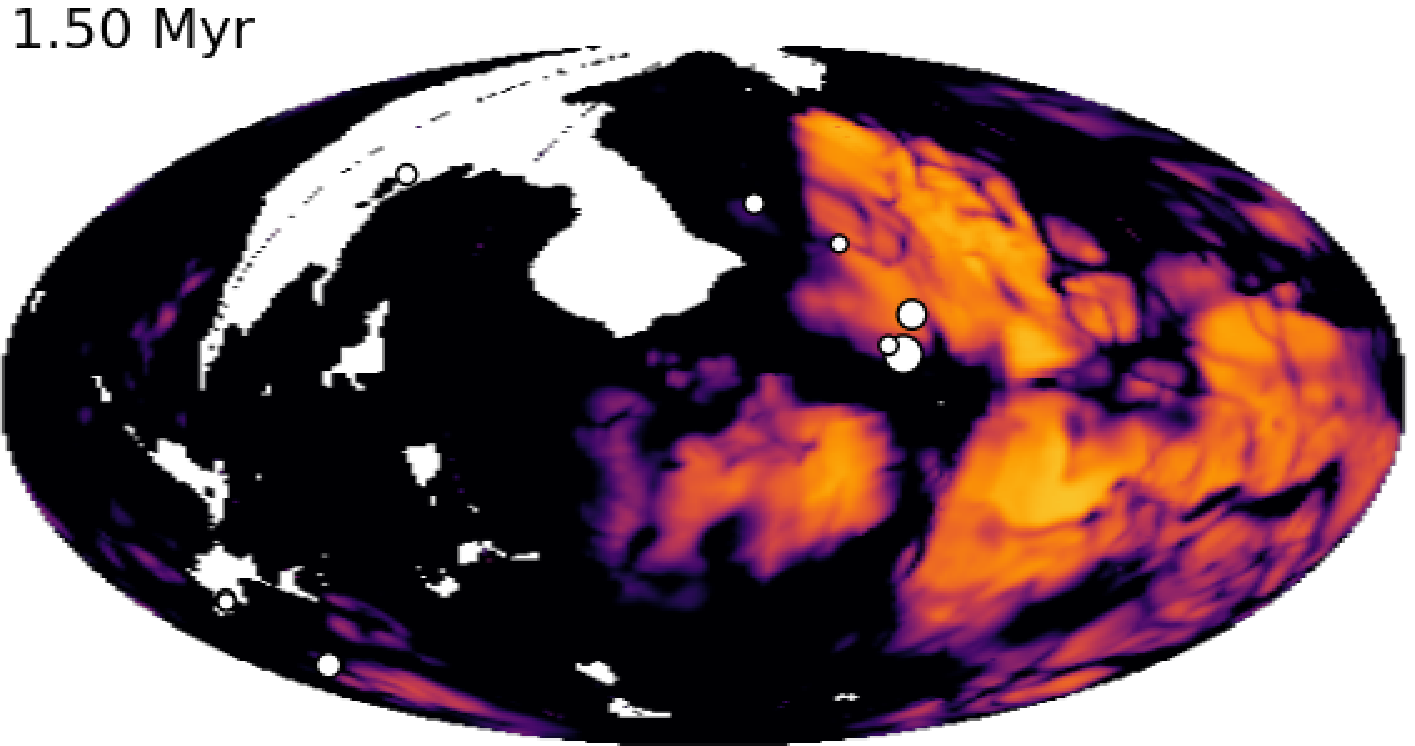} & \includegraphics[width=0.45\textwidth]{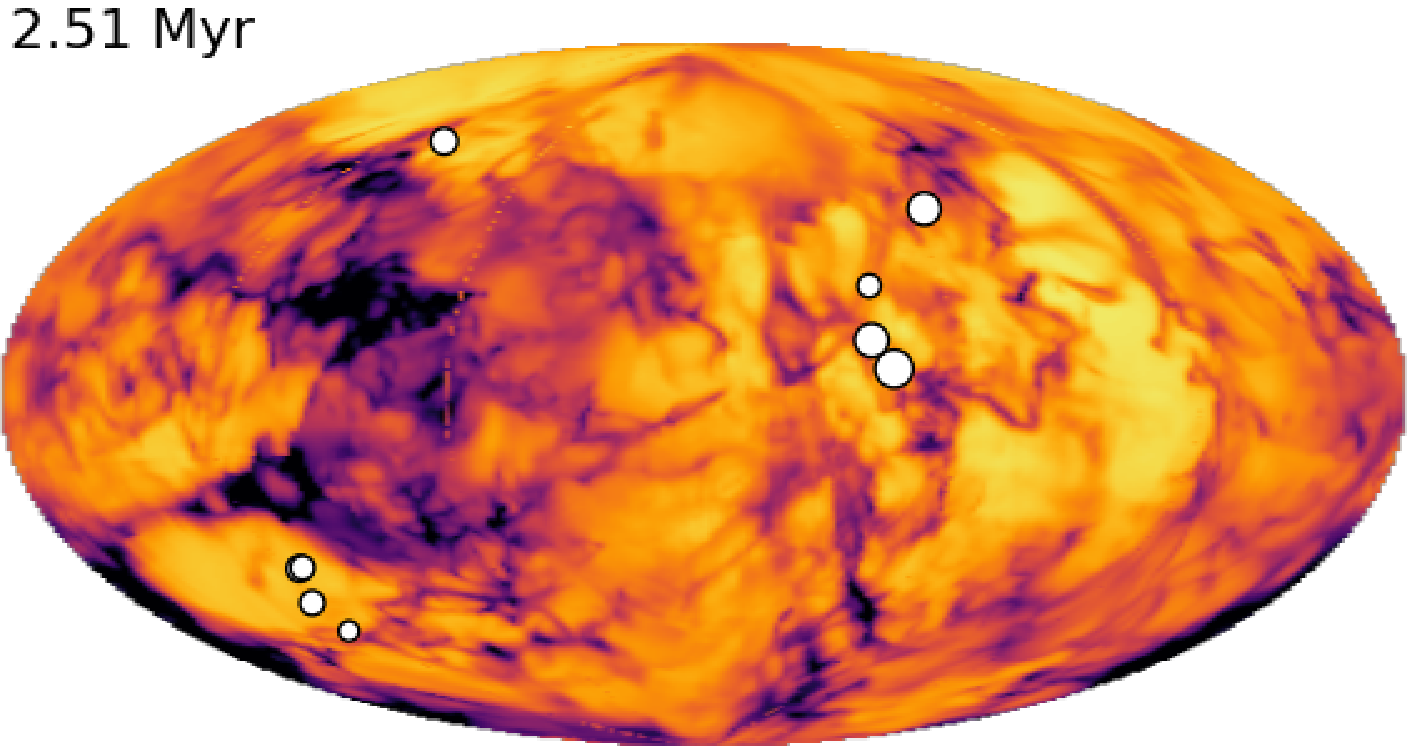} \\
  \includegraphics[width=0.45\textwidth]{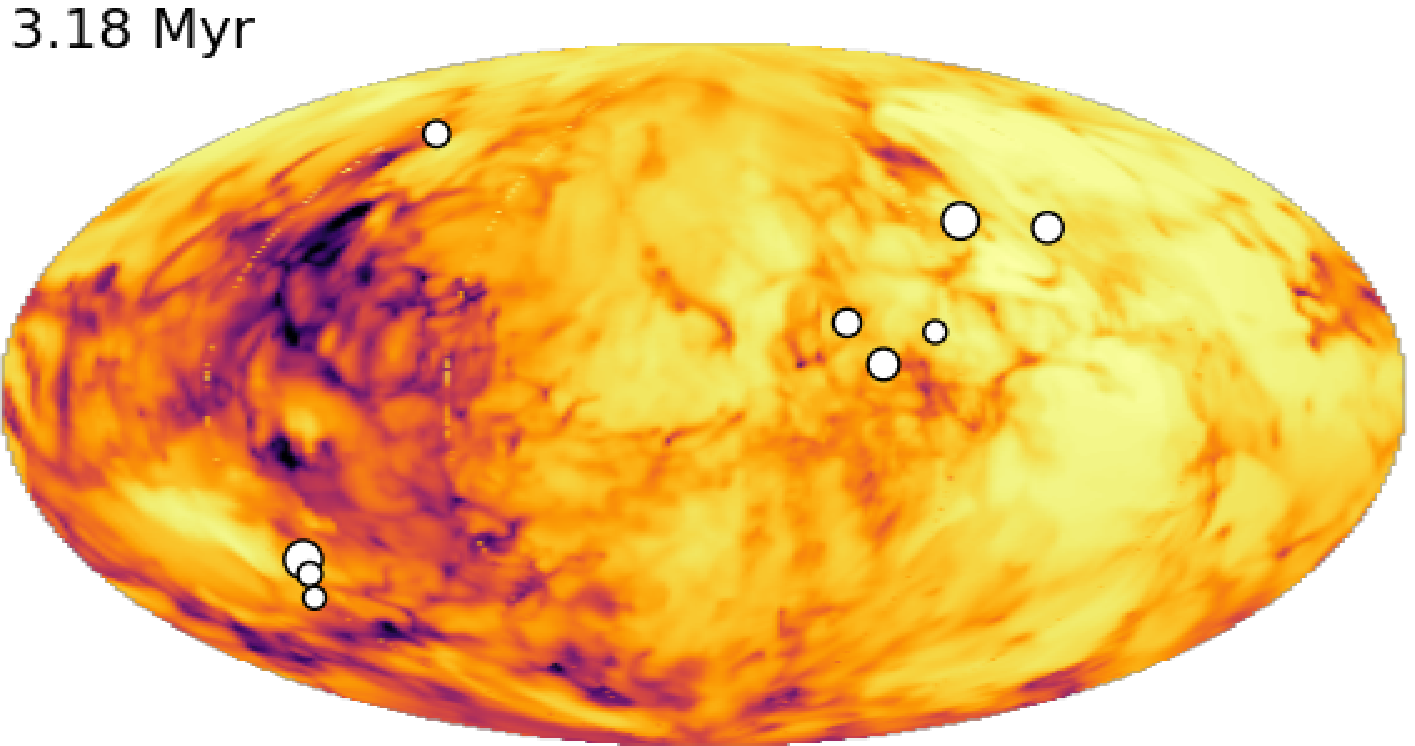} & \includegraphics[width=0.45\textwidth]{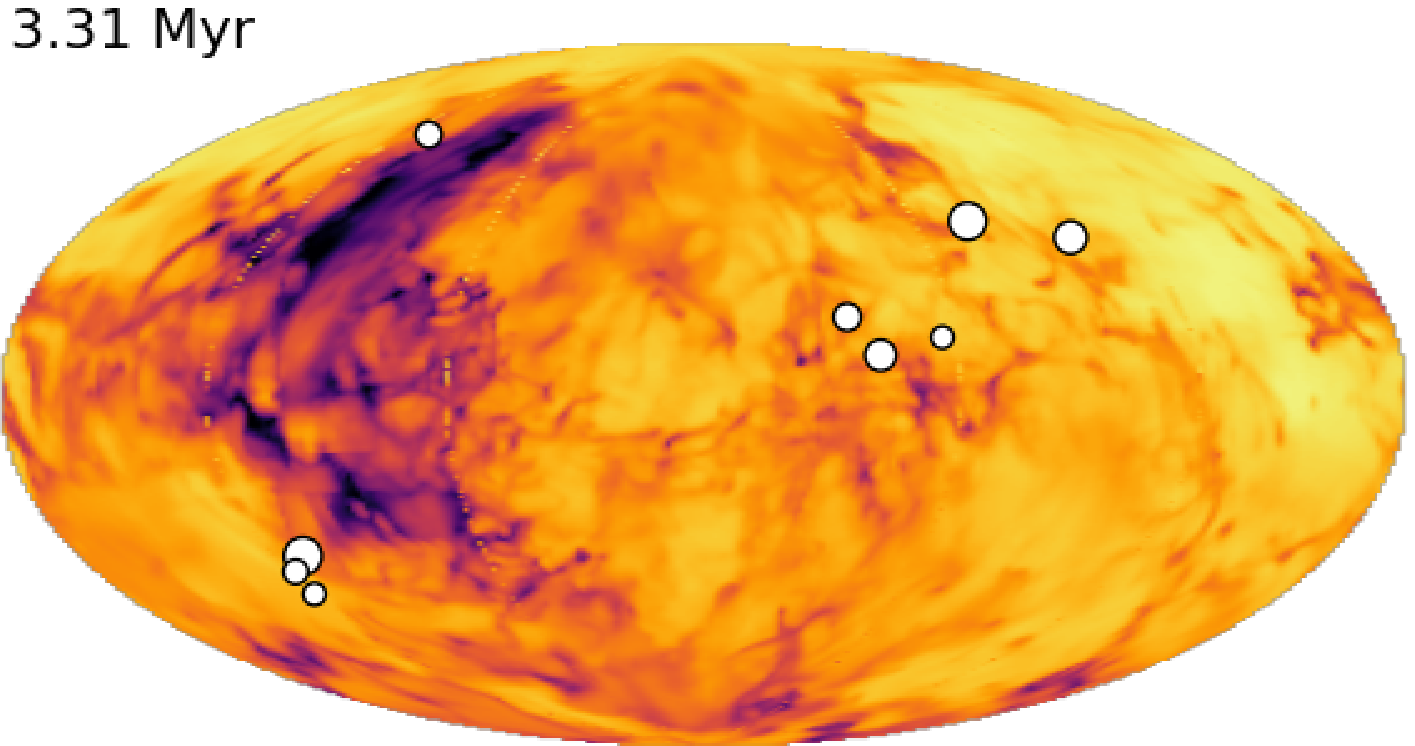} \\
  \includegraphics[width=0.45\textwidth]{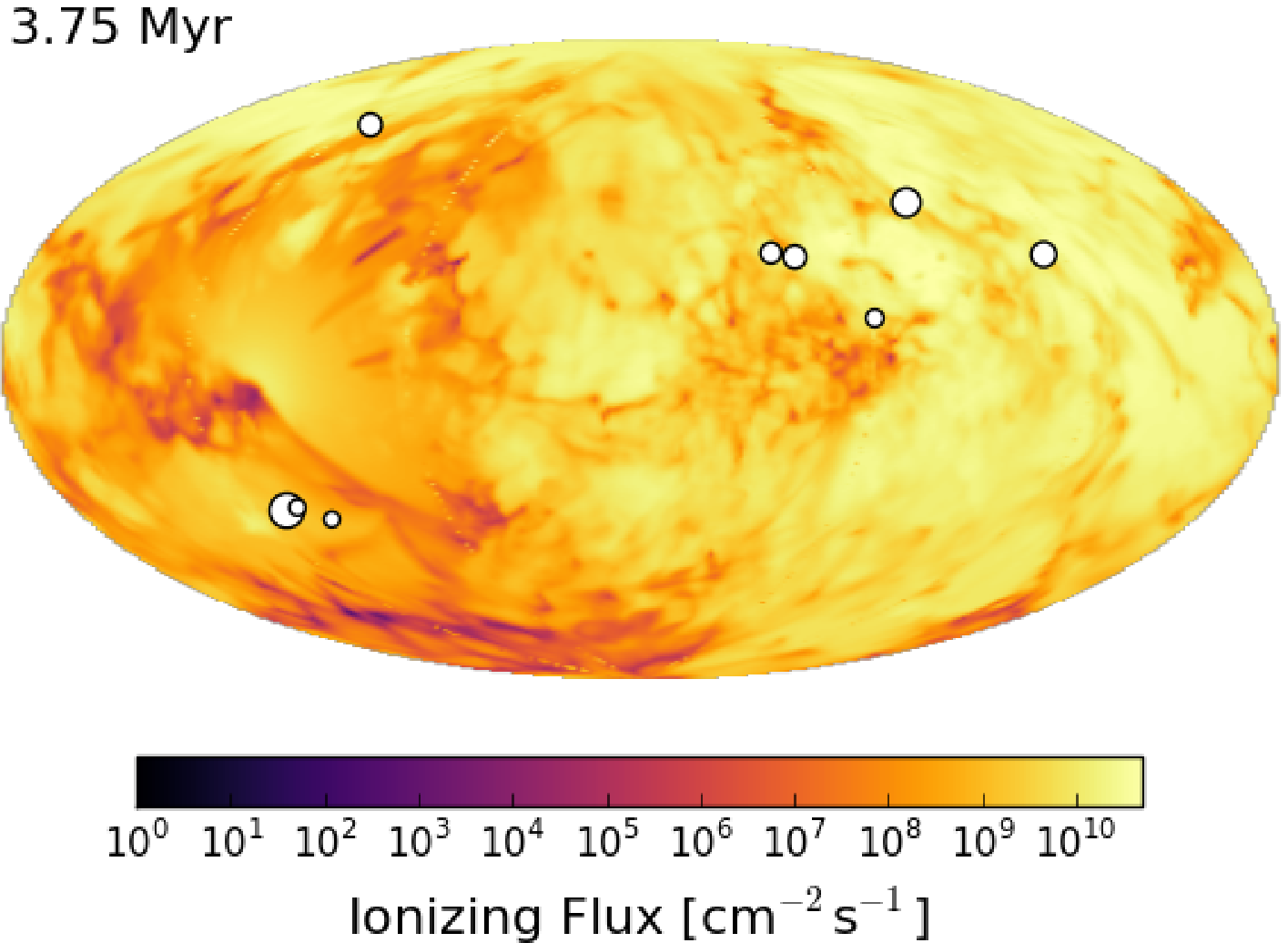} & \includegraphics[width=0.45\textwidth]{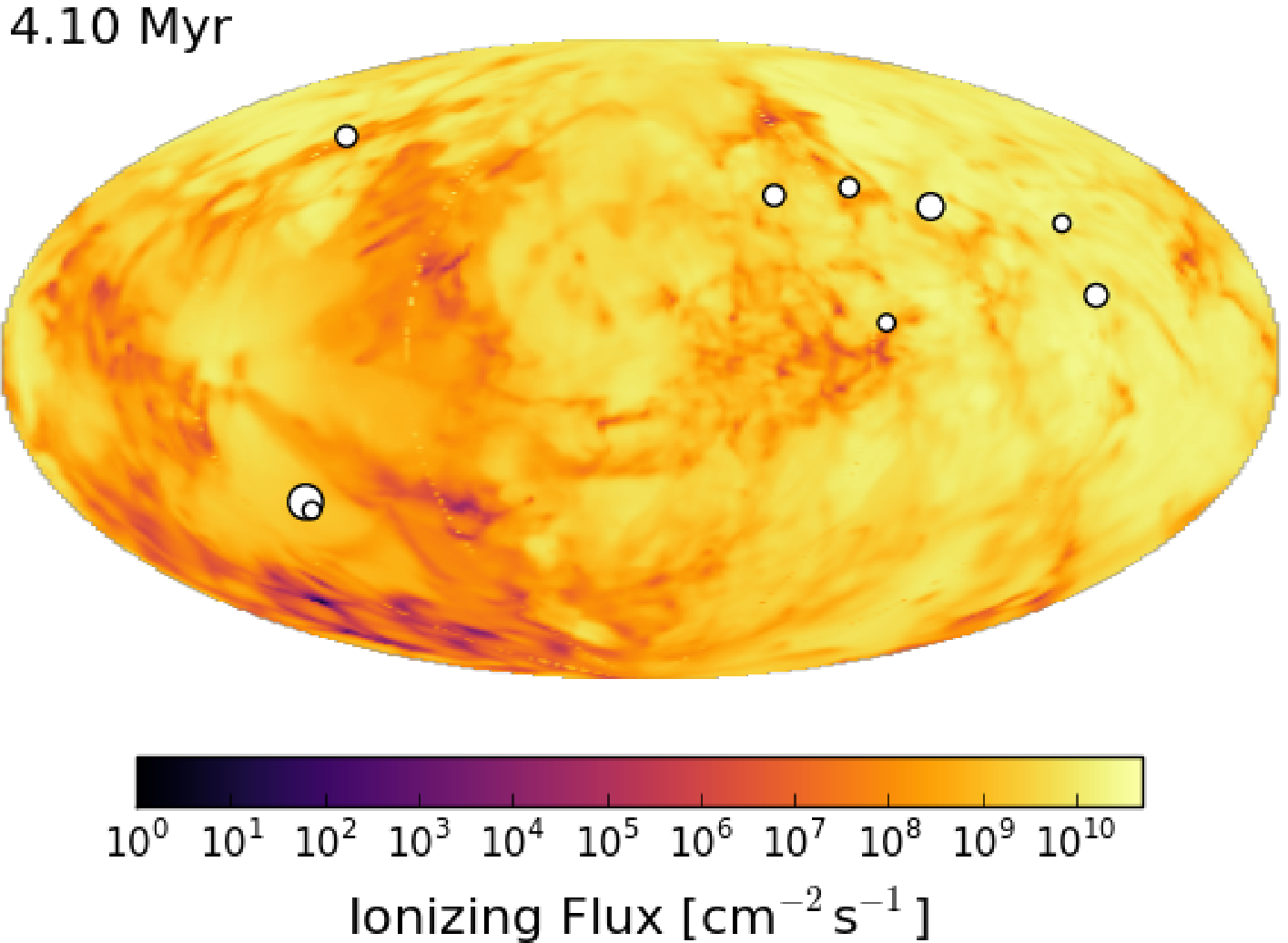} \\
\end{tabular}
\caption{Maps of the ionizing photon flux across a spherical surface of radius 33.8 pc (corresponding to the initial GMC radius) shown at 6 different times. White circles represent the closest location 
of the 10 most luminous clusters to the sphere. More luminous clusters are shown by larger circles. The maps were produced using a Hammer projection, which is an equal area projection.}
\end{figure*}

Radiative transfer is treated via a hybrid-characteristics raytracer developed by \cite{Raytracer} and adapted for astrophysical use by \cite{Peters2010}. This scheme treats
both ionizing and non-ionizing radiation and makes use of the DORIC package \citep{Mellema} to solve the ionization equations. While the DORIC package is capable of treating a large number of species, we consider hydrogen to be the only gas component for simplicity.
The flux of ionizing photons, $F_*$ from an individual source is given by,

\begin{equation}
F_{*} = \frac{S_{*}}{4\pi r^2}e^{-\tau}
\end{equation}

\noindent where $S_*$ is the cluster's ionizing photon rate in s$^{-1}$, $r$ is the distance between the source and cell of interest, and $\tau$ is the intervening optical depth.
The opacity to non-ionizing radiation is represented by the Planck mean opacities from \cite{Pollack}, which are used because the raytracer has no frequency dependence. We adopt a 
single UV opacity in neutral gas of $\kappa$ = 775 cm$^{2}$ s$^{-1}$ from \cite{Li2001}. This opacity is scaled by the neutral fraction of the gas, so completely ionized regions have 
an opacity of zero.

We make use of sink particles \citep{Federrath2010} to model star cluster formation with a custom subgrid model to represent star formation within the clusters \citep{Howard2014}. We adopt
a threshold density for formation of 10$^4$ cm$^{-3}$ which is based on observations of star-forming clumps \citep{Lada2003}. Our subgrid model within cluster sink particles (henceforth referred to as clusters), divides the cluster mass into two types; stars, and the remaining gas mass (denoted 
as the reservoir). We convert the reservoir to stars by randomly distributing the mass into main sequence stars via a \cite{Chabrier2003} IMF with an efficiency of 20\% per freefall time, 
where the freefall time is taken to be 0.36 Myr. The IMF is sampled every tenth of a freefall time to allow cluster propeties to evolve smoothly over time. Newly accreted gas 
is added to the reservoir (ie. gas which available for star formation during the next IMF sampling step). The masses of all stars formed in the cluster are recorded, and analytical fits provided by \cite{Tout} are used to 
determine each star's total and ionizing luminosity. The cluster's luminosity is then the sum of its consituents, which is then used by the raytracer.

\begin{figure*}[!htb]
\centering
\begin{tabular}{ c c }
  \includegraphics[width=0.48\textwidth]{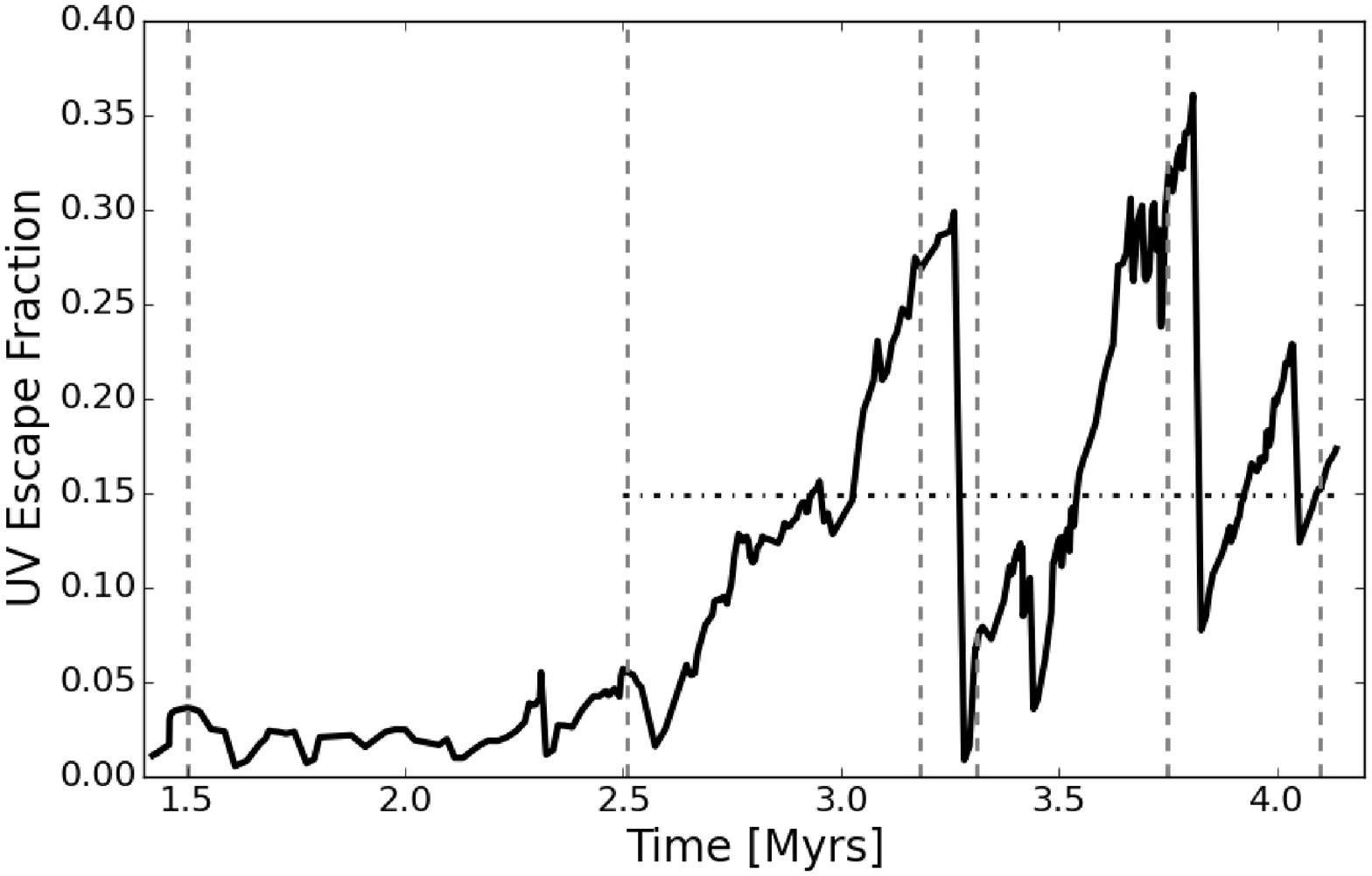} & \includegraphics[width=0.48\textwidth]{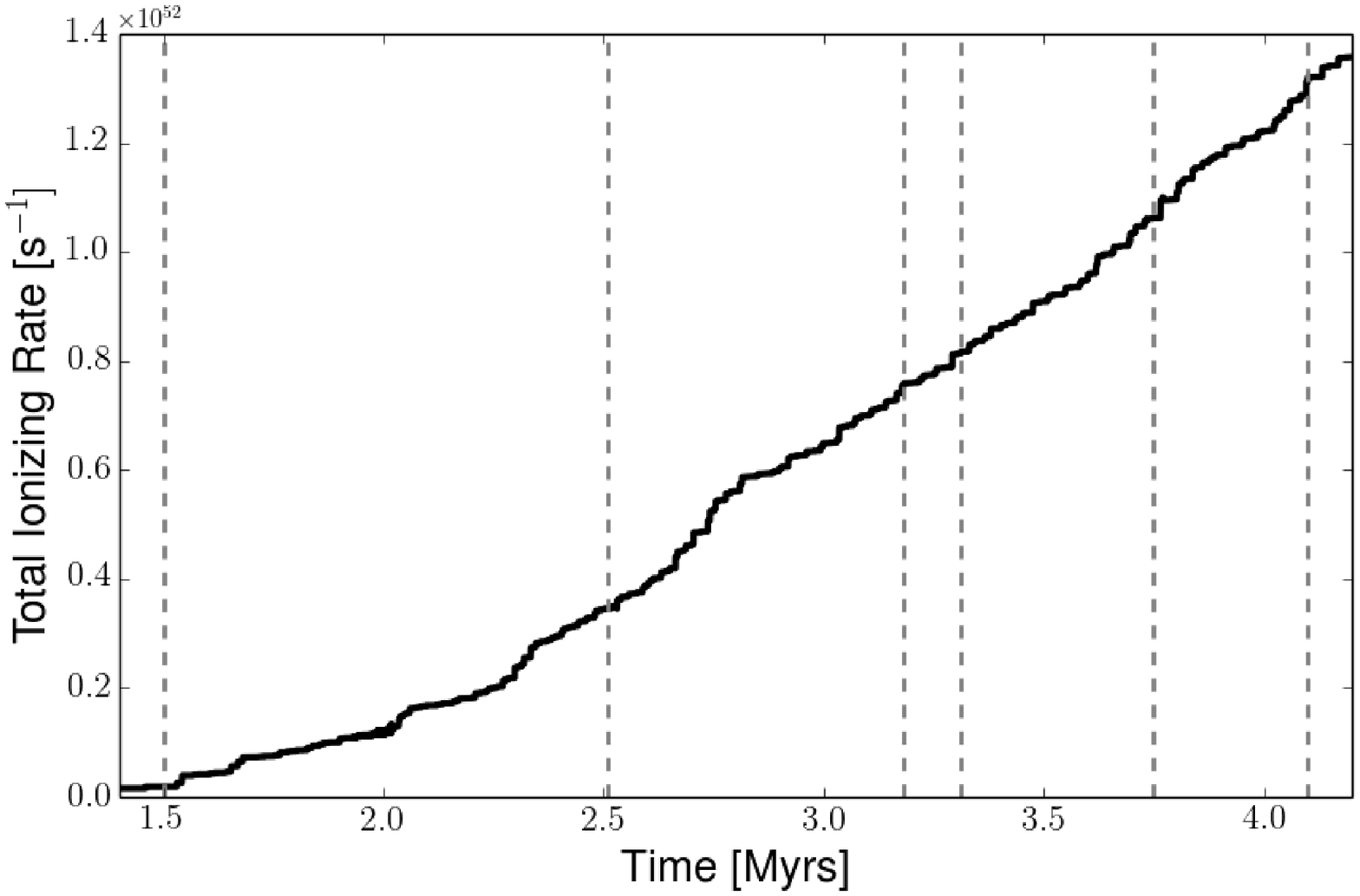} \\
  \includegraphics[width=0.48\textwidth]{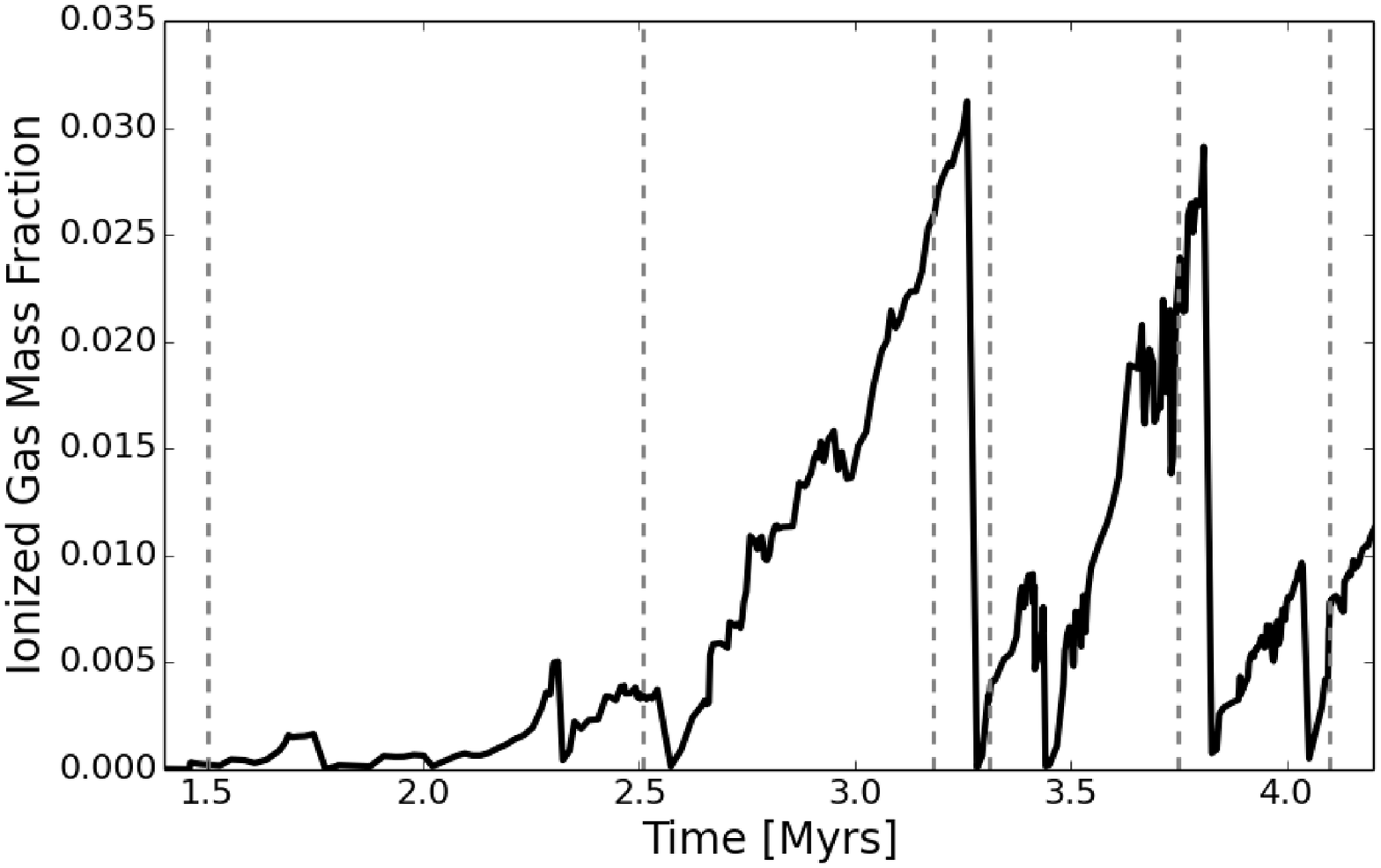} & \includegraphics[width=0.48\textwidth]{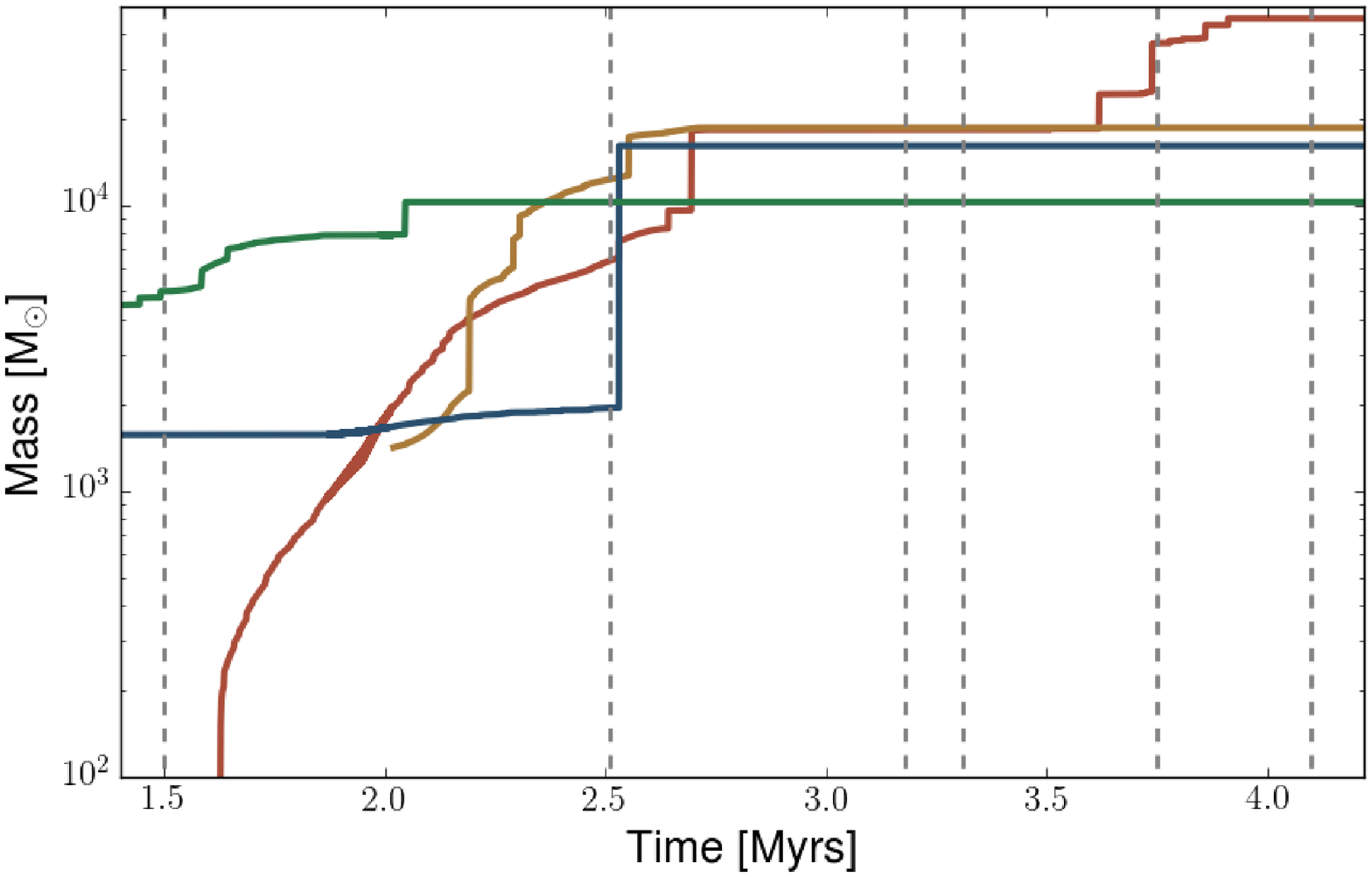} \\
\end{tabular}
\caption{Top Left: The total UV escape fraction across the spherical surface presented in Figure 1. The vertical lines, shown in all panels, correspond to the times shown in Figure 1 (1.50, 2.51, 3.18, 3.31, 3.75, and 4.10 Myr). The horizontal line shows the average f$_{esc}$ from 2.5 to 4.2 Myr.
The escape fraction is defined here as the total number of photons crossing the surface divided by the total number of photons being produced by the clusters. Note that we only include clusters above the mass threshold described in Section 2, since
clusters below this threshold are not included in the radiative transfer calculations. Top Right: The total ionizing photon rate produced by clusters above the mass threshold 
for radiation. Bottom Left: The fraction of gas, by mass, which has an ionization fraction of greater than 95\%. Bottom Right: The mass evolution of the four most massive clusters, shown for reference. Discrete
jumps in mass are due to cluster merging events. The complete mass evolution tracks can be found in \cite{Howard2016}.}
\end{figure*}

In order to reduce the computational time, we apply a mass threshold of 1000 M$_{\odot}$ in \emph{stars} (which typically have $\sim$1 O star), below which clusters do not radiate. Clusters below the threshold
continue to accrete gas and form new stars but they are not included in the radiative transfer calculation. 

\section{Results} \label{sec:results}

To study the spatial distribution of the escaping UV flux from the cloud, we produce maps of the ionizing flux across a spherical surface which are presented in Figure 1. The radius of this sphere corresponds to the intial cloud radius
of 33.8 pc and all clusters are contained within the surface. The maps were made using a Hammer projection which was chosen because it is an equal area 
projection. We also include the locations of the 10 most luminous clusters (accounting for 93\% of the final ionizing luminosity), projected to the closest location on the sphere, in white circles. Note that the clusters
are not actually located on this spherical surface, but are contained within its volume. 

The first panel, plotted at 1.5 Myr, shows the ionizing flux shortly after the first clusters begin to radiate. A large fraction of the surface is not receiving any 
ionizing photons, shown by the white patches. This is because at this time, the clusters have only recently formed (meaning that their total 
ionizing luminosity is low compared to their final values). 

In the same panel, the regions that are receiving ionizing photons are concentrated in the upper right quadrant. Note that the most luminous clusters appear in a grouping 
towards the right side as well, suggesting that these clusters are responsible for much of the emission observed outside the cloud. There is also some flux associated with the cluster in the
bottom left quadrant of this panel. 

At 2.5 Myr, we see that the entire surface is now being traversed by UV photons from the clusters. The flux of photons, however, is not spatially uniform. 
Since the flux on the sphere's surface depends on the intervening column density, the presence of dense clumps and filaments manifests itself as regions with lower flux. We note
that the simulation has virialized ($\alpha$ $=$ 1) at 2.5 Myr, so any further turbulence is driven by gravitational collapse of the gas (see \cite{Howard2016} for details and \cite{Klessen2010} for a more general discussion of accretion driven turbulence). 

As the total ionizing luminosity increases and the total mass in gas decreases, the presence of these dark filaments becomes less pronounced. At 3.18 Myr, 
only the left side of Figure 1 shows regions with low flux. The grouping of clusters on the right of this Figure is likely responsible for the higher 
flux in that region. From 3.75 Myr onwards, the flux is more spatially uniform due to increased cluster luminosities and lower total gas mass. 

The above visualizations show that the ionizing flux can vary significantly over both space and time within a GMC. 

In Figure 2, we focus on the evolution of f$_{esc}$ from the cloud. We define f$_{esc}$ as the total number of photons crossing the spherical surface previously discussed in Figure 1, divided by the summed total of all photons being generated by the clusters. 
Its time evolution is shown in the top left panel of Figure 2. Note that we only include the clusters which are above the mass threshold discussed in Section 2, since these are the clusters that are used by the radiative transfer scheme.

\begin{figure*}
\centering
\begin{tabular}{ c c c}
  \includegraphics[width=0.32\textwidth]{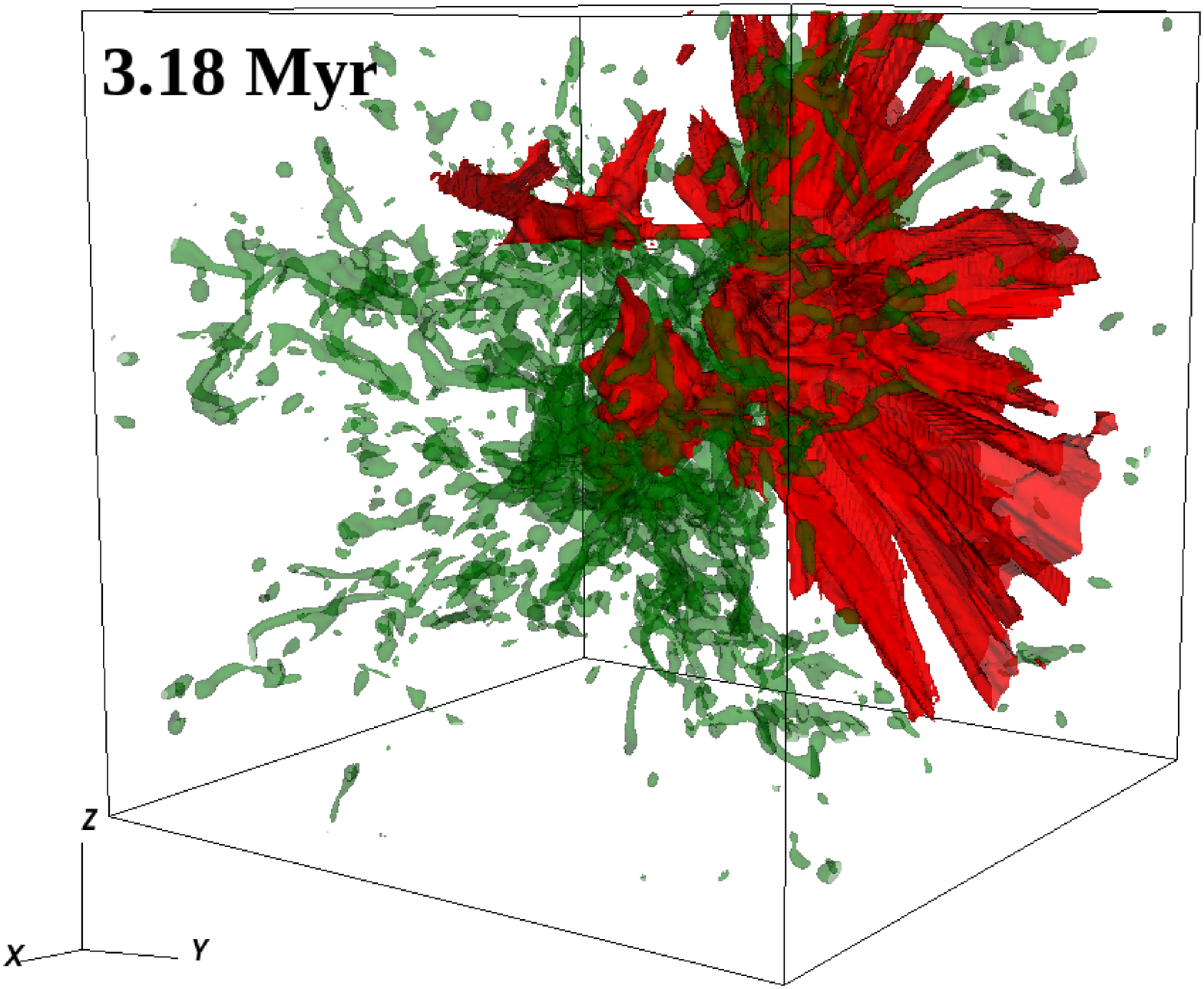} & \includegraphics[width=0.32\textwidth]{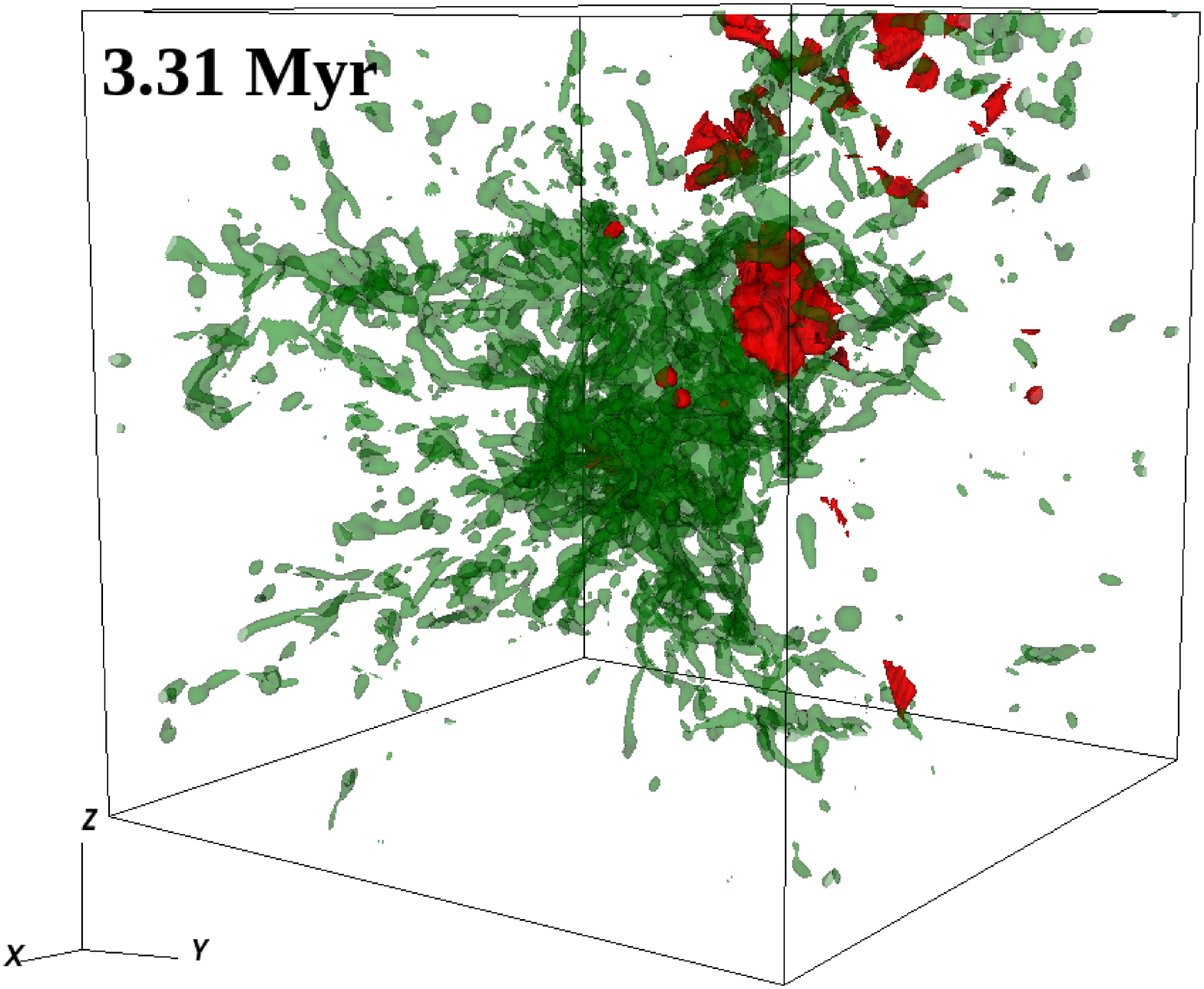} & \includegraphics[width=0.32\textwidth]{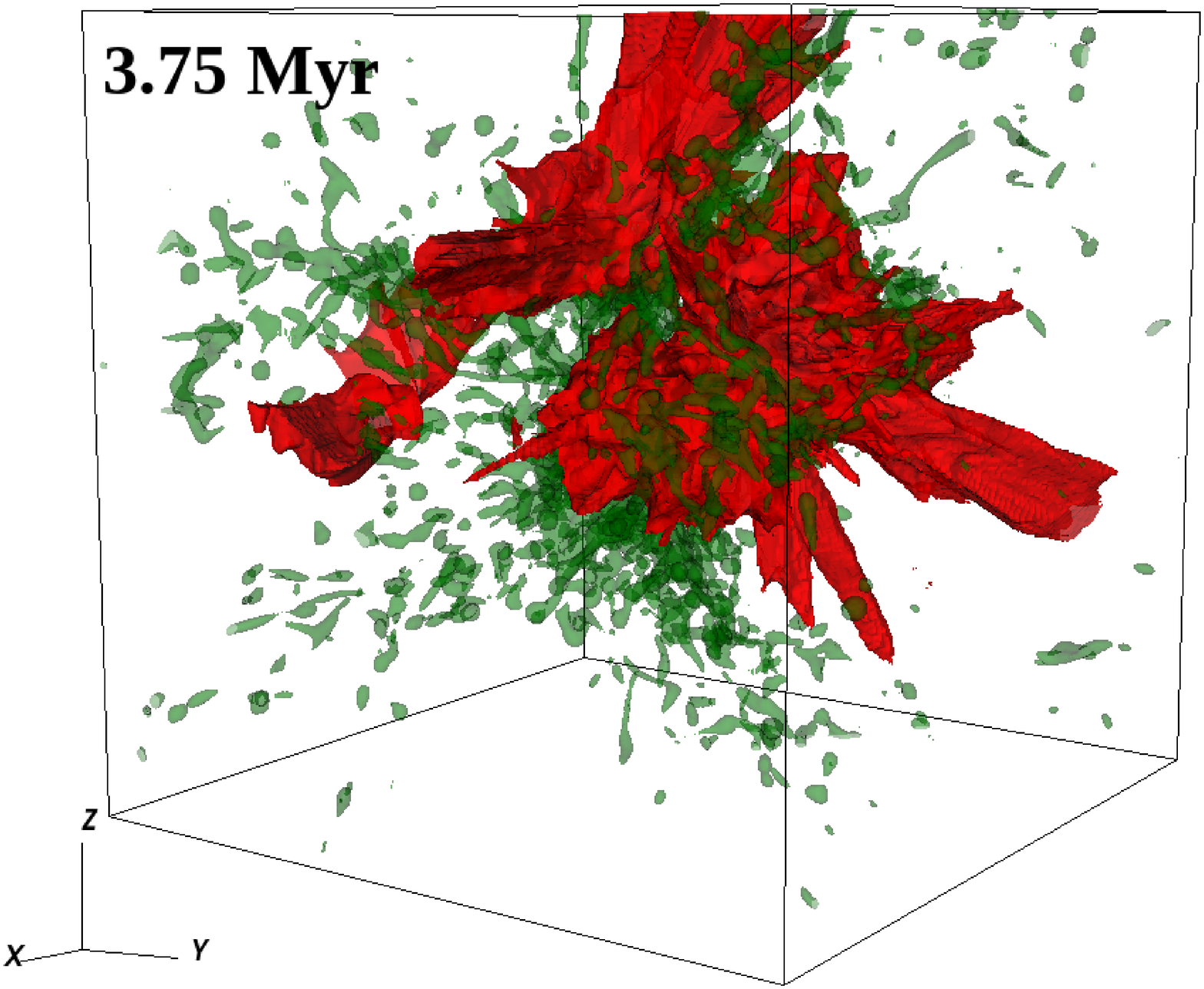} \\
\end{tabular}
\caption{3d images of the density (shown in green) and ionized regions (shown in red) at 3.18, 3.31, and 3.75 Myr from left to right, respectively. These images correspond to the first peak in f$_{esc}$ in Figure 2, the trough at 3.31 Myr, and the second peak at 3.75 Myr. The density contours represent densities of $\sim$30 cm$^{-3}$ and the box side length is
80 pc.}
\end{figure*}

The total escape fraction remains low at approximately 3\% between 1.5 and 2.5 Myr. After 2.5 Myr, f$_{esc}$ rises to a peak of 30\% at 3.25 Myr, followed
by a sudden drop. The escape fraction begins to rise again, reaching a peak of 37\% at 3.8 Myr. The average f$_{esc}$ from the first rise at 2.5 Myr to the end of the simulation, 
shown by the horizontal line, is 15\%.

The rising f$_{esc}$ and subsequent rapid drops are not due to changes in the ionizing photon output from the clusters, which is shown in the top right panel of Figure 2. These
clusters are accreting new gas vigourously from their surroundings and building new, massive stars as time progresses. The increase in the ionizing photon output is steady and shows no 
distinct features which correspond to the features seen in f$_{esc}$. 

Rather, the ionization structure of the gas is responsible for the variable f$_{esc}$. In the bottom left panel of Figure 2, we plot the fraction of gas mass which has an ionization 
fraction of greater than 95\%. This Figure clearly mirrors the features seen in f$_{esc}$, with an increasing f$_{esc}$ corresponding to an increase in the mass
fraction of ionized gas. Recall that UV opacity in ionized regions is significantly lower than in neutral regions. 

While the change in the ionized mass fraction is low, peaking at $\sim$3\%, the HII regions can spatially occupy a significant fraction of the simulation volume, typically 
filling large voids that are interspersed between dense filaments. Since clusters are the source of radiation and, therefore, tend to exist in HII regions, photons
can travel large distances due to the reduced opacity resulting in higher photon fluxes near the boundary of the simulation volume. 

However, the size and shape of HII regions is not constant. Both observations \citep{DePree2013,DePree2015} and simulations \citep{Peters2010, UCHII, Madrid2011,Klassen2012-1} show that 
the size of an HII region can flucuate on short timescales, a phenomenon described as "flickering". The dynamic and anisotropic nature of the gas, in combination with
dynamic clusters, can result in HII regions becoming shielded to radiation due to changes in density between the source and the ionized regions. The formerly irradiated gas 
then recombines, causing the HII to flicker.

\begin{figure*}
\centering
\begin{tabular}{ c c c}
  \includegraphics[width=0.48\textwidth]{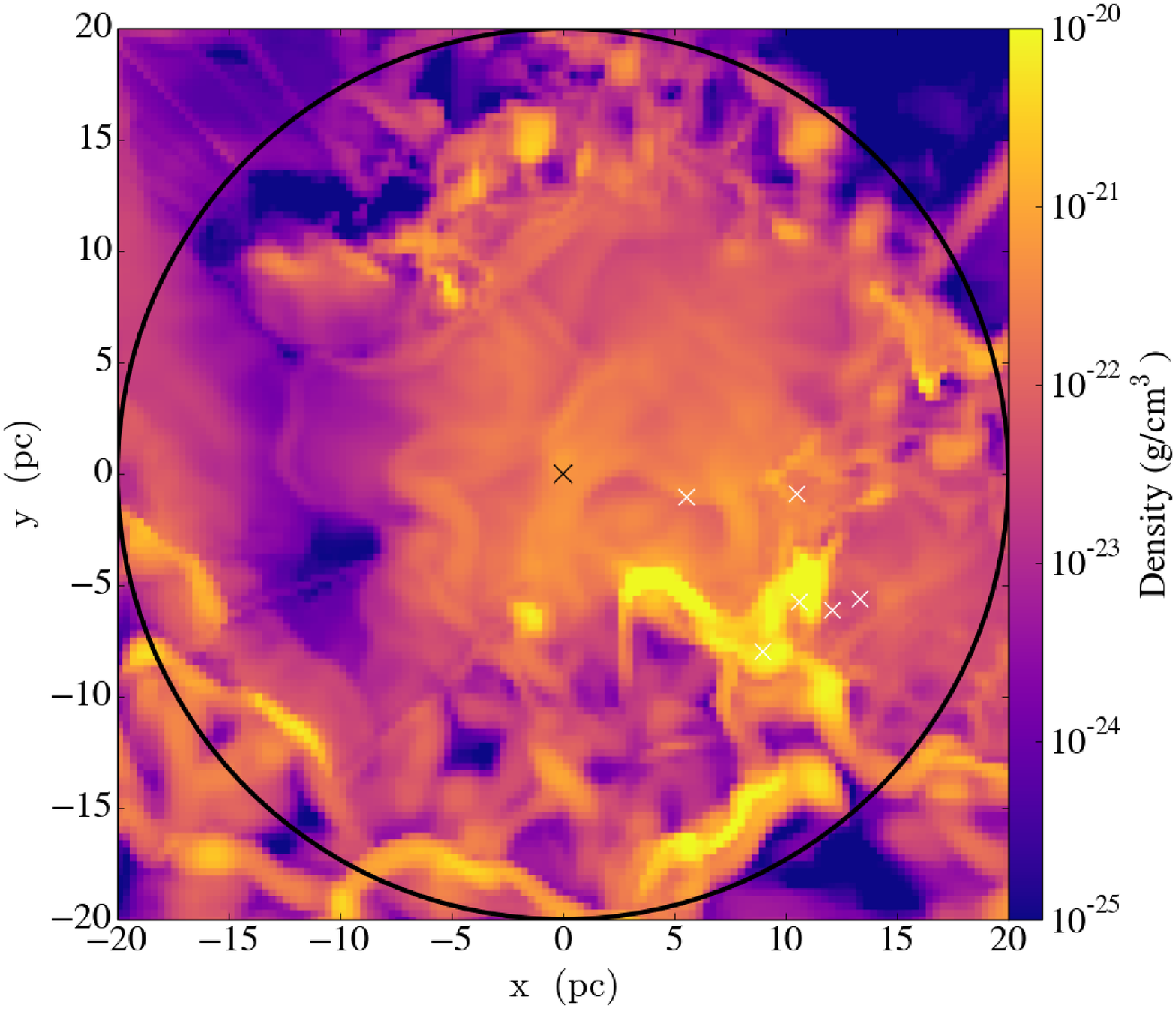} & \includegraphics[width=0.48\textwidth]{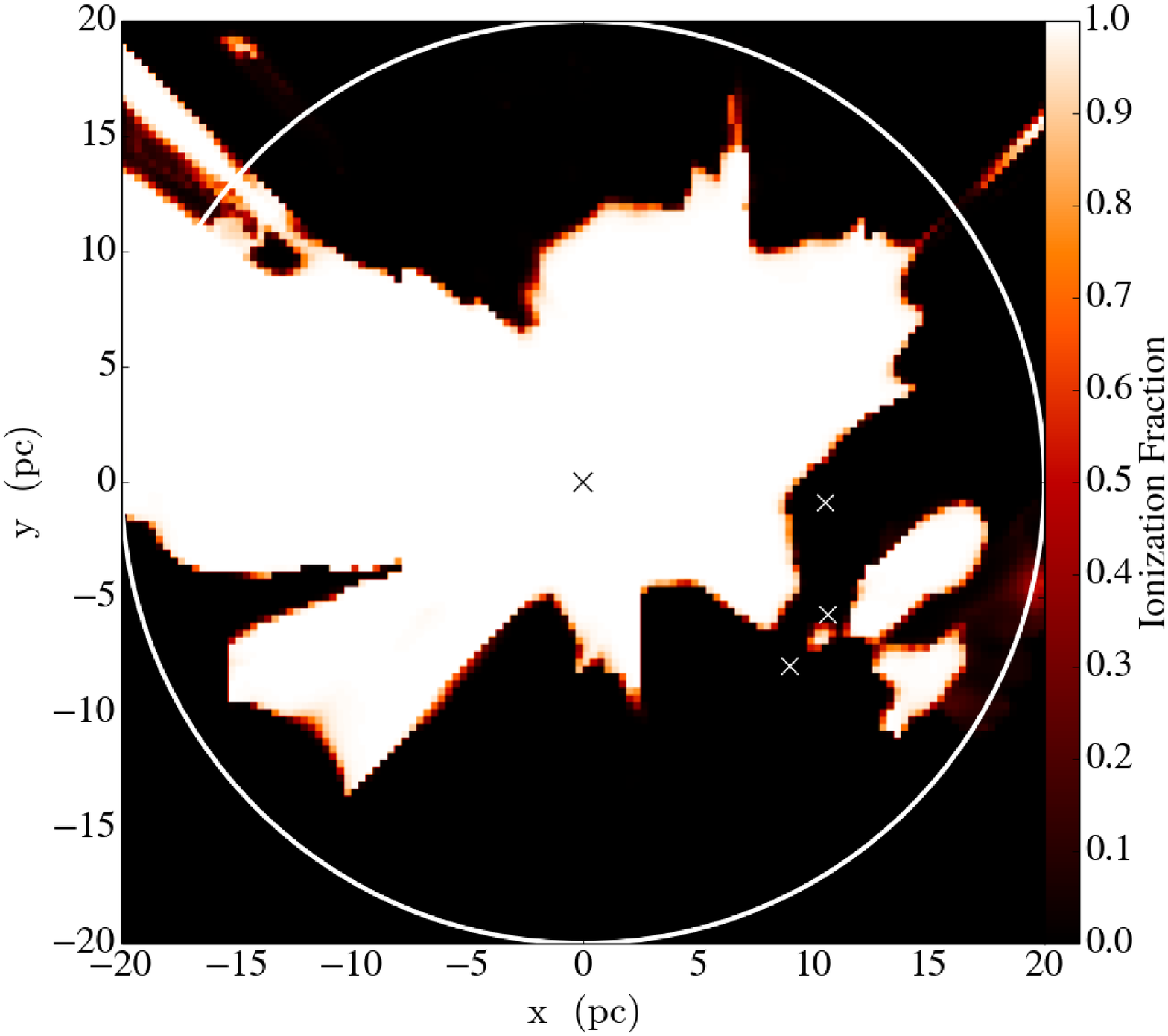} \\
  \includegraphics[width=0.48\textwidth]{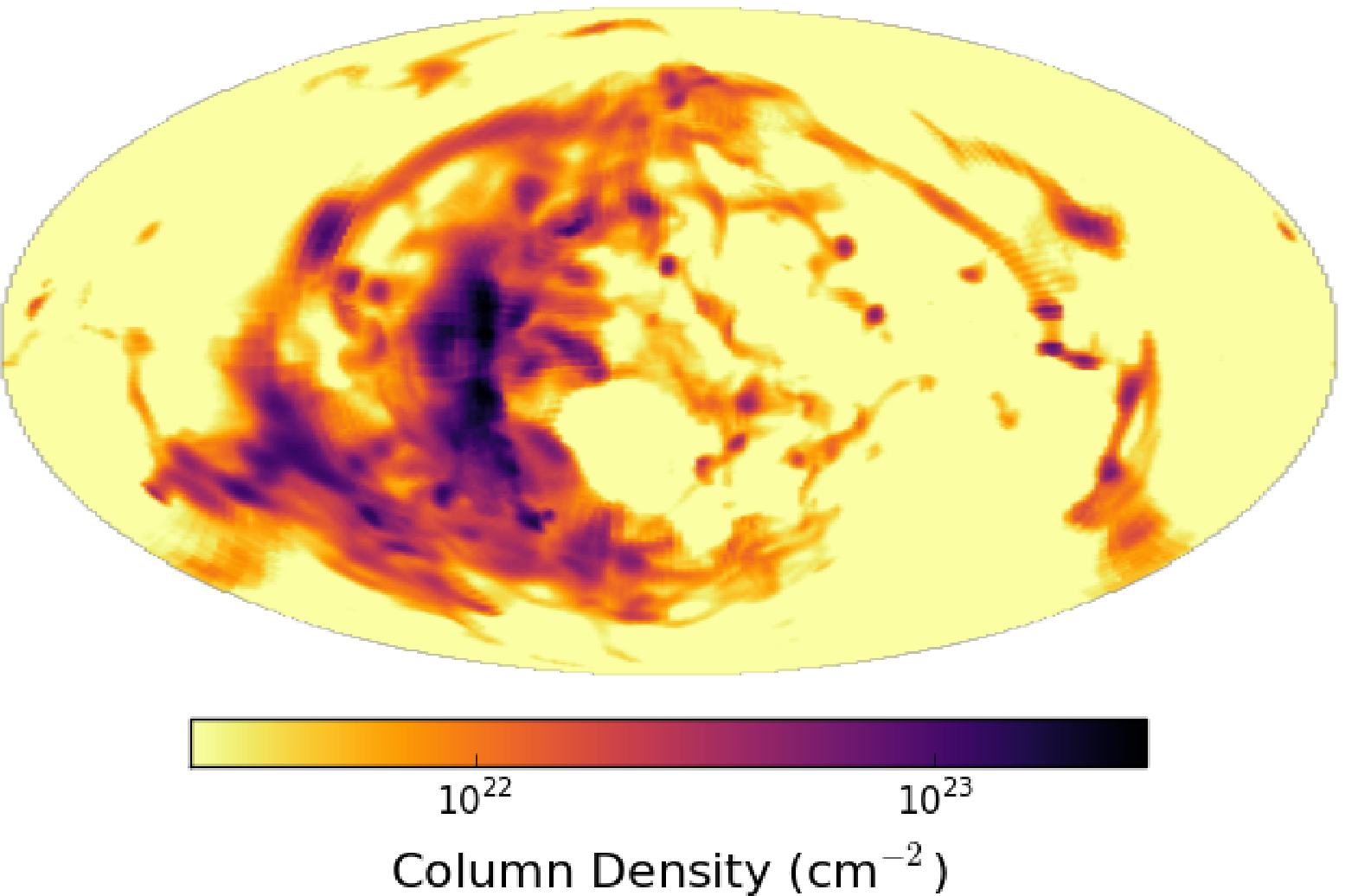} & \includegraphics[width=0.48\textwidth]{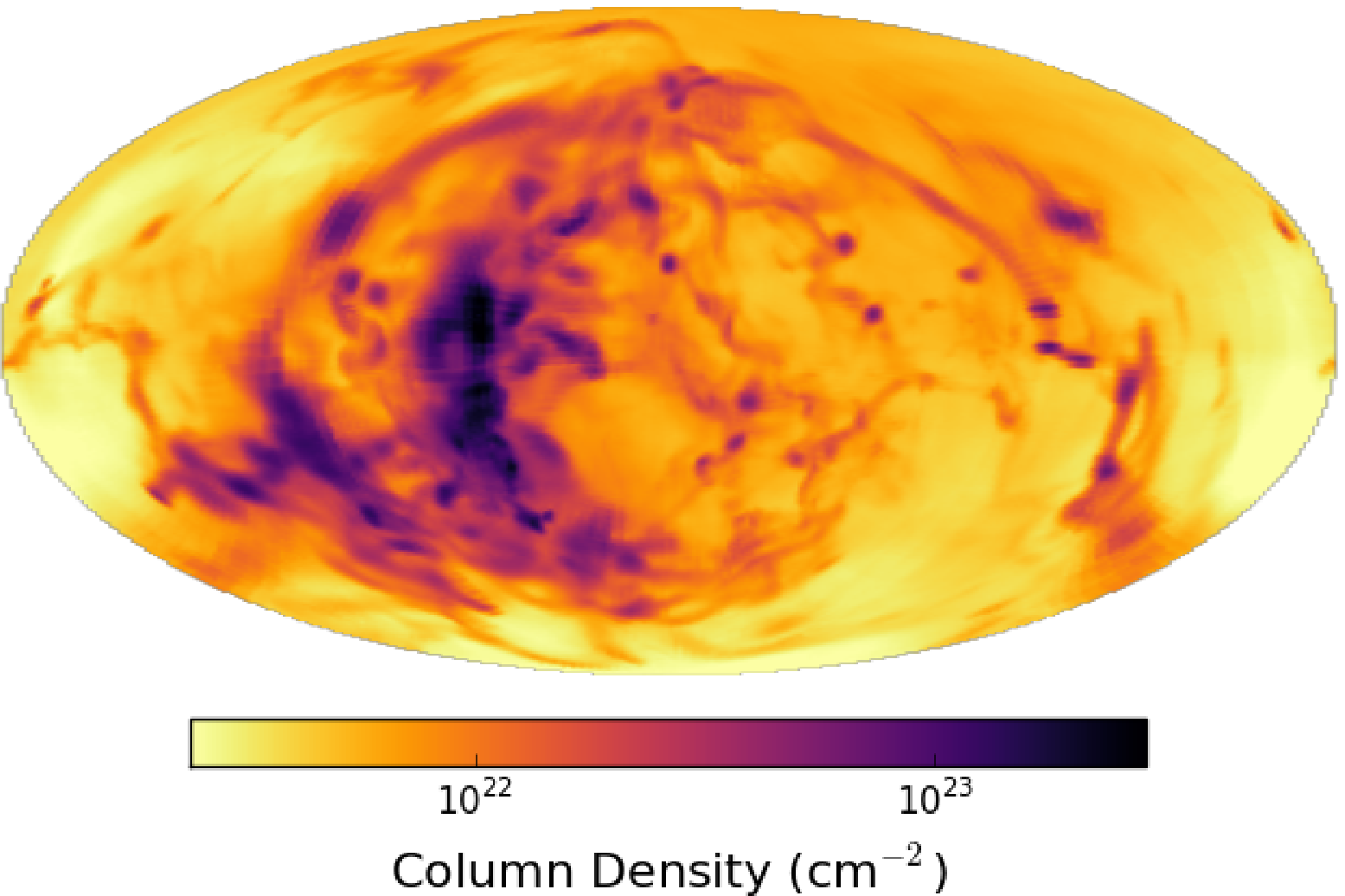} 
\end{tabular}
\caption{Top Left: A slice of density centered on the location of the most luminous cluster associated with the collapsing HII region at 3.25 Myr. Top Right: The same slice in the previous
panel except the ionization fraction of the gas is displayed. Bottom Left: A Hammer projection of the neutral gas column density at a 20 pc spherical surface centered on the same cluster, plotted before the HII region collapses at 3.25 Myr.
Bottom Right: The same but $\sim$35,000 yr after the Bottom Left panel. The HII region has collapsed by this time.}
\end{figure*}

A visual demonstration of this flickering is displayed in Figure 3, which shows 3-dimensional images of density, in green, and HII regions, in red. The green density contours show gas at
$\sim$100 cm$^{-3}$ which is the typical density of the filaments out of which the clusters form. The entire simulation volume is shown and the box side length is 83 pc.

The left most panel of Figure 3 shows the state of the simulation at 3.18 Myr, corresponding to the first pronounced peak in f$_{esc}$. A large HII region has developed on one
side of the cloud which extends away from the dense, central gas to the boundary of the simulation volume. The middle panel, shown at 3.31 Myr, shows the decrease in the size of the HII region
which is responsible for the deep trough in f$_{esc}$ at 3.25 Myr. The right most panel of Figure 3, plotted at the second peak of f$_{esc}$ at 3.75 Myr, shows
that the HII region has expanded again to a similar size as seen in the first panel.

To investigate the cause of the variable HII region size, we focus our analysis on one luminous cluster which is associated with the HII region. This cluster is the second most luminous in the simulation with a final ionizing luminosity of 1.40$\times$10$^{51}$ s$^{-1}$. The most luminous cluster was not chosen because it is deeply embedded in the dense, central gas and therefore its associated HII region is small in comparison to the one which extends to the boundary of the simulation volume, as seen in Figure 3.

We drew lines of sight which originate at the cluster's position 
and extend a distance of 20 pc through the large HII region. This was done at two times, one just before the HII region collapses for the first time (at 3.25 Myr) and one immediately after the collapse ($\sim$35,000 yr after the first image). We can then examine how the density and the recombination rate differ before and after the HII region collapse along these lines of sight.

We find that the radiative recombination rate along the lines of sight increases significantly immediately after the HII region collapses, increasing from $\sim$5$\times$10$^{-8}$ to 1$\times$10$^{-6}$ cm$^{-3}$ s$^{-1}$. The radiative recombination rate is given by $\alpha n^2$, assuming an ionization degree of 100\%, where $\alpha$ is the radiative recombination coefficient and $n^2$ is the square of the number density. The recombination coefficient varies with temperature as, 

\begin{equation}
\alpha = 2.59\times 10^{-13}\left(\frac{T}{10^4 K}\right)^{-0.7}
\end{equation}

\noindent where $T$ is the gas temperature in Kelvin.

The radiative recombination rate increases after the collapse for two reasons. Firstly, the density immediately surrounding the cluster increases, likely due to the turbulent nature of the surrounding gas. Secondly, as the region cools, the recombination coefficient increases. 

We also examined the quantity $\alpha x^2 n^2$, where $x$ is the ionization fraction of the gas, which removes the assumption of a 100\% ionization fraction. In this case, we see the opposite trend and the recombination rate
drops from 5$\times$10$^{-8}$ to 2$\times$10$^{-13}$ cm$^{-3}$ s$^{-1}$. Despite the increase in density and the recombination coefficient, the recombination rates decrease significantly due
to the low ionization fraction of the gas after the HII region collapses. 

We find that $n^2$ increases by a factor of $\sim$1.5 - 4 along the lines of sight within a 1 pc radius of the clusters location. This increased density limits the amount of radiation that propagates to larger radii. The gas can then recombine and cool from $\sim$10$^{4}$ K, typical of HII regions, to $\sim$10 K which is the temperature floor adopted in the simulation.

This can be visualized by examining the neutral column density from the luminous cluster through the HII region, as shown in Figure 4. The top panels show a slice of density (left) and 
ionization fraction (right) centered on the luminous cluster associated with the HII region which collapses at 3.25 Myr. These images are plotted before the HII region collapses. At this time, the cluster is no longer deeply embedded in the massive cold filament out of which it formed in the first place.
The bottom panels of Figure 4 shows a Hammer projection of the neutral gas column density across a spherical surface
of radius 20 pc centered the same cluster before the HII region collapse (left) and after the collapse (right). A 20 pc radius circle is shown in the density and ionization fraction
slices for reference. The column density projections clearly show that the region which was previously ionized has increased in column density after the HII region collapses.

The increase in density surrounding the massive cluster can be understood through turbulent shocks in the surrounding ionized gas. We measured the local gas velocity dispersion at the location
of the cluster immediately before the HII region collapse to be 10.1 km s$^{-1}$ (corresponding to a Mach number of 1.14). Thus a density fluctuation can cross the cluster's radius of 0.78 pc in $\sim$76,000 years. This is comparable to the $\sim$35,000 years
it takes for the HII region to collapse. A passing shock could therefore lead to a local density enchancement causing the HII region to collapse in the observed time. As the gas recombines, it cools and shields regions further along the line of sight.

\begin{figure*}
\centering
\begin{tabular}{c c}
  \includegraphics[width=0.49\textwidth]{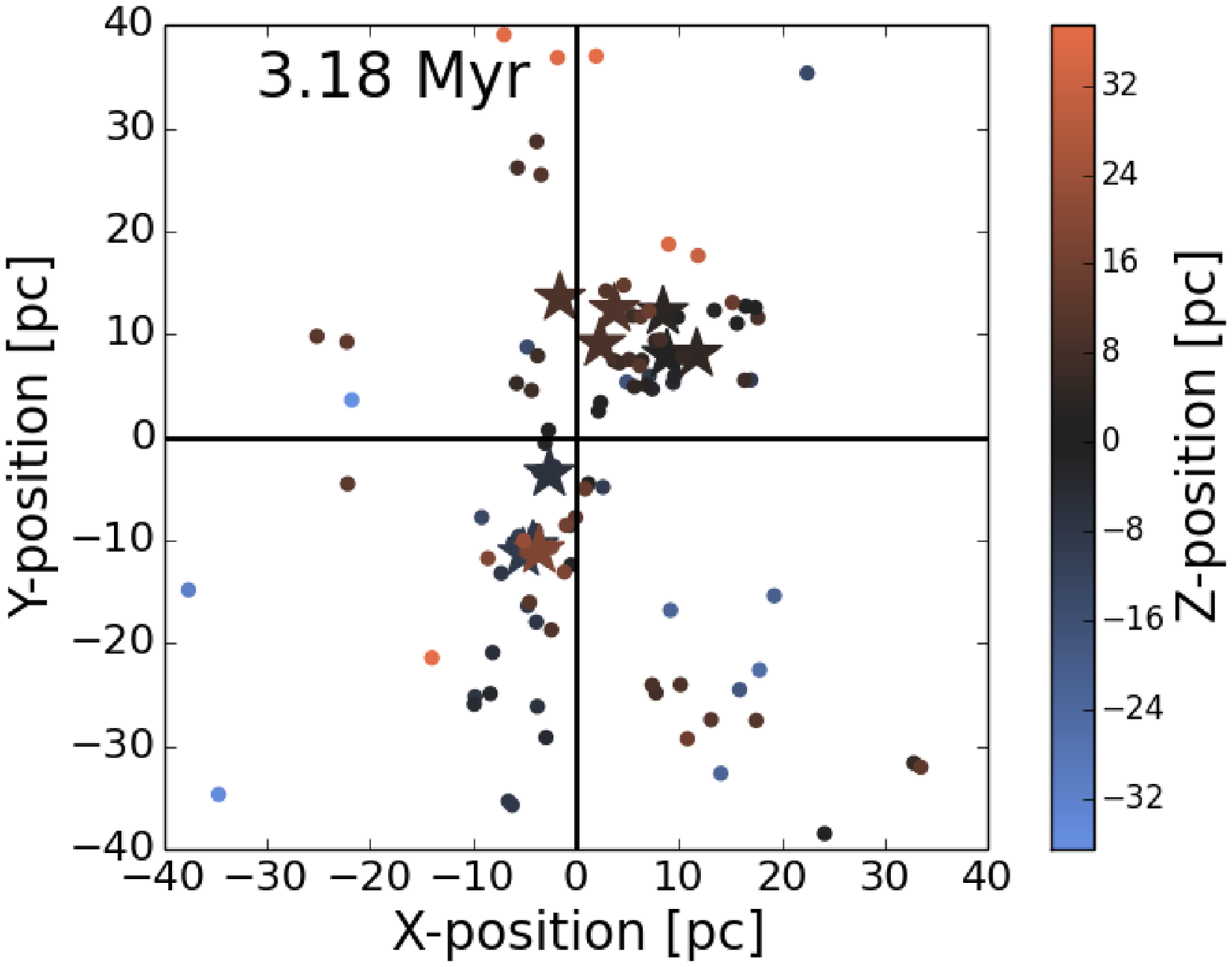} & \includegraphics[width=0.49\textwidth]{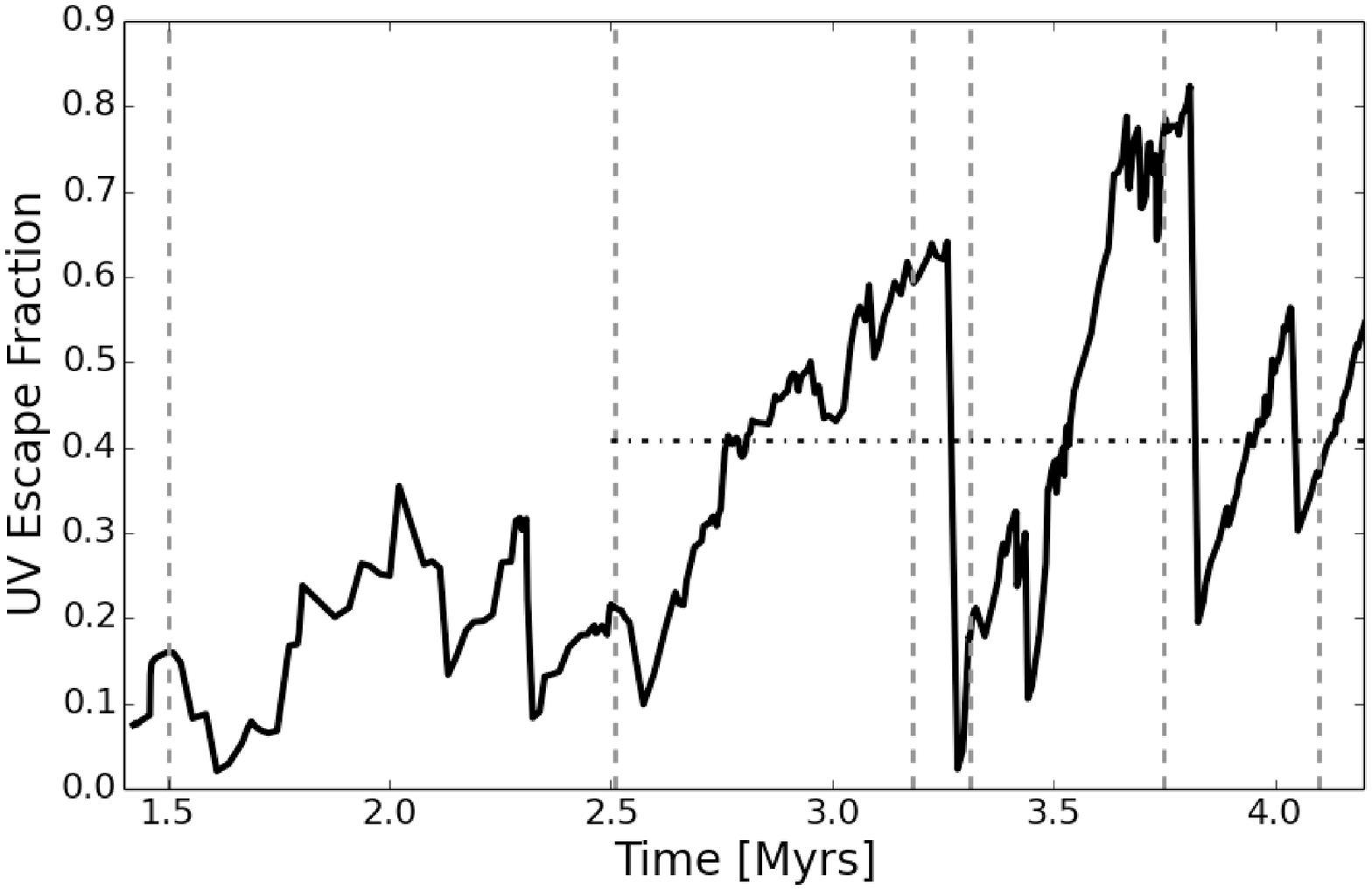} \\
\end{tabular}
\caption{Left: The positions of all clusters plotted at 3.18 Myr. The stars represent the locations of the 10 most massive clusters and colours represent Z-position. Right:
The escape fraction, f$_{esc}$, across a spherical surface of radius 25 pc. This is the smallest possible radius which contains 90\% of the total ionizing luminosity within its volume
and therefore minimizes the contribution to the flux across the surface from outside sources.}
\end{figure*}

This further analysis supports our claim
that the dynamic nature of the gas, which at this point in time causes the density to increase within the HII region, is responsible for the strong fluctuations in the size of the HII
regions being produced by luminous clusters, and hence in the value of f$_{esc}$.

The above discussion has focused on f$_{esc}$ from the entire molecular cloud which is a useful quantity when studying the build-up of the ISRF or estimating the global escape fraction from a galaxy as a whole. We may instead investigate the escape
fraction from smaller regions surrounding luminous clusters to follow the evolution of f$_{esc}$ as a function of distance from the cluster. We are limited in this regard due 
to the fact that the raytracer used to compute the radiative transfer only tracks the total flux in each grid cell with no directional information about the incoming rays. This means
that if we calculate the flux across a small spherical surface centered on a luminous cluster, it will likely include contributions from sources outside the sphere.

Still, we can calculate f$_{esc}$ across a surface if the majority of the total ionizing luminosity is being generated within its volume. This minimizes the contribution to
the total flux from outside sources. We find that the 10 most luminous clusters generate 90\% of the total ionizing luminosity and are located a maximum of 24.3 pc from the simulation
center. We therefore repeat the f$_{esc}$ calculation for a sphere of radius 25 pc instead of 33.8 pc, which was presented in Figure 2.

For reference, we show a 2-dimensional projection of the position of all clusters in the left panel of Figure 5. The 10 most luminous clusters are shown by the stars, and all clusters
are colored by their Z-position in the cloud. We only show the positions at one time, 3.18 Myr which corresponds to the first peak of f$_{esc}$ in Figure 2, to illustrate that the 
clusters are not strongly grouped together but instead cover the cloud's entire extent.

The right panel of Figure 5 shows f$_{esc}$ across a spherical surface of radius 25 pc. Comparing to the top left panel of Figure 2, we see that f$_{esc}$ across the smaller 
surface well within the cloud has a similar temporal evolution with pronounced peaks at 3.25 and 3.75 Myr, but the values are $\sim$2 times larger. The average f$_{esc}$ from 2.5 to 4.2 Myr is 41\%, compared
to 15\% for ionizing radiation that escapes from the surface of the cloud. The early evolution (ie. less than 2.5 Myr) of f$_{esc}$ is also significantly enhanced, likely due to the generation of small HII regions surrounding 
the luminous clusters which are not large enough to extend to the edge of the simulation volume.

Overall, these results suggest that f$_{esc}$ decreases with cloud radius as one moves through the cloud and out its surface. This trend has been found observationally by \citet{Pelle2015} who noted that the global f$_{esc}$ from the LMC and SMC is estimated to be 4\% and 11\% respectively, while f$_{esc}$ from individual star-forming regions can be as high as $\sim$60\%.

The values presented in Figure 2 are particularly important for the global ISRF and ISM structure since they represent the escape fraction from the surface of an entire 10$^6$ M$_{\odot}$ GMC. Clouds of this mass are host to the most massive stellar clusters which dominate the stellar feedback and overall luminosity of a galaxy \citep{Harris1994,MacLow2004,McKee2007,Klessen2016}.

\section{Discussion and Conclusions} \label{sec:conc}

We computed the UV photon escape fraction from a turbulent, 10$^6$ M$_{\odot}$ GMC using the astrophysical code FLASH. The cloud, taken from \cite{Howard2016}, is initially unbound with a virial parameter of 3 and sink particles are used to model the formation of star clusters. Our simulations end just before supernovae explosions could disrupt the star clusters and remove surrounding gas.

Our analysis indicates that the flux is both highly anisotropic due to the filamentary and clumpy nature of the intervening turbulent gas as well as highly variable in time. As time progresses, the flux naturally increases since the clusters contain more massive stars.

The integrated escape fraction, defined as the total number of photons leaving the cloud divided by the total number of photons being produced by all clusters, also varies significantly over time. For the first 2.5 Myr of evolution, f$_{esc}$ remains low at $\sim$5\%. There are two distinct peaks in the escape fraction at 3.25 and 3.8 Myr, with a maximum escape fraction of 30\% and 37\% at the two peaks. The average f$_{esc}$ from the onset of large HII regions at 2.5 Myr to the end of the simulation is 15\%. 
The average f$_{esc}$ increases to 41\% if we instead consider a smaller surface well inside the cloud at a radius of radius 25 pc, as compared to the GMC's surface which is at a radius of 33.8 pc.

The peaks of f$_{esc}$, and subsequent troughs, are tied to the local gas density structure surrounding the luminous clusters which determines the size of the HII region they produce. At a peak, the HII is large and extends towards the boundary of the simulation volume. At a trough, the HII region is only a small fraction of its previous size. The collapse of the HII region is tied to the turbulent nature of the gas surrounding the luminous clusters which causes the density to vary with time. As it increases, so does the recombination rates and the HII regions shrinks.

We argue that calculations of the photon escape fraction on galactic scales
require knowledge of f$_{esc}$ for individual GMCs and for the star forming
complexes in their interior. For many applications (eg. cosmic reionization), this is computationally prohibitive
and we suggest to use an average value of f$_{esc}$ = 15\% for a 10$^{6}$ M$_{\odot}$ GMC with fluctuations
of a factor of two superimposed on timescales of about 1 Myr. 


\acknowledgments

We thank Bill Harris and Eric Pelligrini for interesting discussions.

C.S.H. and R.E.P. thank the the Zentrum f\"ur Astronomie der Universit\"at Heidelberg (ZAH), Institut f\"ur Theoretische Astrophysik
(ITA), as well as the Max-Planck-Institut f\"ur Astronomie (MPIA), for their generous support during R.E.P's sabbatical leave (2015/16)
 and C.S.H.'s extended visit (Oct. - Nov.,2015).  

R.E.P. is supported by Discovery Grants from the Natural Sciences and Engineering Research Council (NSERC) of Canada. C.S.H. acknowledges financial support provided
 by the Natural Sciences and Engineering Research Council (NSERC) through a Postgraduate scholarship. The FLASH code was in part developed by the DOE
supported Alliances Center for Astrophysical Thermonuclear Flashes (ASCI) at the University of Chicago. This work
was made possible by the facilities of the Shared Hierarchical Academic Research Computing Network (SHARCNET:www.sharcnet.ca) 
and Compute/Calcul Canada.

R.S.K. acknowledges support from the European Research Council under the
European Community’s Seventh Framework Programme (FP7/2007-2013) via the ERC
Advanced Grant 'STARLIGHT: Formation of the First Stars' (project number 339177). R.S.K.
further thanks for funding from the Deutsche Forschungsgemeinschaft (DFG) in the
Collaborative Research Center SFB 881 ‘The Milky Way System’ (subprojects B1, B2, and
B8) and in the Priority Program SPP 1573 ‘Physics of the Interstellar Medium’ (grant numbers
KL 1358/18.1, KL 1358/19.2).

\allauthors

\listofchanges

\end{document}